# The value of big data for analyzing growth dynamics of technology-based new ventures

*(accepted manuscript)*

Maksim Malyy[a], Zeljko Tekic[a,b], Tatiana Podladchikova[a]

[a]Skolkovo Institute of Science and Technology, Bolshoy Boulevard 30/1, New Campus Skoltech, 143026 Moscow, Russia

[b]HSE University, Graduate School of Business, Shabolovka 26-28, 119049 Moscow, Russia

E-mail: maksim.malyy@skoltech.ru (M. Malyy), z.tekic@skoltech.ru (Z. Tekic), t.podladchikova@skoltech.ru (T. Podladchikova)

## Abstract

This study demonstrates that web-search traffic information, in particular, Google Trends data, is a credible novel source of high-quality and easy-to-access data for analyzing technology-based new ventures (TBNVs) growth trajectories. Utilizing the diverse sample of 241 US-based TBNVs, we comparatively analyze the relationship between companies' evolution curves represented by search activity on the one hand and by valuations achieved through rounds of venture investments on another. The results suggest that TBNV's growth dynamics are positively and strongly correlated with its web search traffic across the sample. This correlation is more robust when a company is a) more successful (in terms of valuation achieved) – especially if it is a "unicorn"; b) consumer-oriented (i.e., b2c); and 3) develops products in the form of a digital platform. Further analysis based on fuzzy-set Qualitative Comparative Analysis (fsQCA) shows that for the most successful companies ("unicorns") and consumer-oriented digital platforms (i.e., b2c digital platform companies) proposed approach may be extremely reliable, while for other high-growth TBNVs it is useful for analyzing their growth dynamics, albeit to a more limited degree. The proposed methodological approach opens a wide range of possibilities for analyzing, researching and predicting the growth of recently formed growth-oriented companies, in practice and academia.

## 1. Introduction

Startups and high potential technology-based new ventures (TBNVs[a]) that emerge from them are considered to be key drivers of economic development, innovation, and job creation on the national and

---

[a] *High potential* underlines that companies focus on and ability to achieve annual sales of over $10M during the first several years of existence (Roure and Keeley, 1990). In the case of a "startup," which is known as the early stage of a TBNV (Kazanjian, 1988; Lee and Lee, 2006), the *high potential* quality is driven together by scalability and repeatability features (Blank, 2013; Kollmann et al., 2016). *Technology-based new ventures* itself represent the firms "recently established by a group of entrepreneurs, based on the exploitation of an invention or technological innovation and which employ a high proportion of qualified employees" (Campos et al., 2011). These companies are understood to be in the focus of venture capitalists and "face unusual time pressures and uncertainty" (Roure and

global levels (Acs and Armington, 2006; Henrekson and Johansson, 2010; Kane, 2010; Mason and Brown, 2014). Bringing novel solutions to existing and emerging problems, startups create new value for their customers, at the same time increasing competition within the economy. More than that – startups are the main agents of disruption. Over the last 15 years, digital-native startups have been scaling globally from zero to billion dollars in value in just a couple of years, changing how things are done across industries and setting standards for the next generation of products and companies (Tekic and Koroteev, 2019).

However, although startups are very important for the economy, and although they attract a lot of interest from researchers, policy-makers, nascent entrepreneurs, and managers alike, analyzing their growth and performance is a challenging task. Despite high and continuous research efforts (for in depth reviews see, for example: Gilbert et al., 2006; Shepherd and Wiklund, 2009; Weinzimmer et al., 1998), our theoretical understanding of how (new) firms grow is limited and develops slowly (Coad, 2007; Gilbert et al., 2006; McKelvie and Wiklund, 2010). One of major reasons for this is the lack of appropriate indicator(s) that will effectively capture growth (Weinzimmer et al., 1998) and help in adequately answering question "how companies grow?" (McKelvie and Wiklund, 2010).

Different methods and data sources have been used in the literature to measure the growth of new ventures. Three the most frequent are sales growth, employee growth, and market share growth (Colombo and Grilli, 2010; Davila et al., 2003; Gilbert et al., 2006). However, these proxies have particular limitations (Shepherd and Wiklund, 2009), which are amplified in the case of startups, temporary organizations (Blank, 2013) that develop new products under conditions of extreme uncertainty (Ries, 2011), and have no or very short operating and performance history. To resolve these issues, scholars recently started to use a company valuation, achieved through funding rounds, as a proxy growth measure (Chang, 2004; Gornall and Strebulaev, 2017; Malyy et al., 2019). However, this approach has the same significant limitation as earlier mentioned approaches – data frequently are non-disclosed. Even when the startup phase ends and a company starts to scale-up, the data scarcity issue does not disappear.

The information availability problem is solved through direct communication with venture founders and top managers. Data are collected either through online surveys (Audretsch et al., 2012; Lee et al., 2001; Zhou et al., 2014), or face-to-face interviews with co-founders and higher management (Bocken, 2015; Carter et al., 1996; Delmar and Davidsson, 2000; Malyy and Tekic, 2018). Online surveys offer access to valuable data but suffer from sample bias due to the low response rates (10-15% in the best cases) and other selection criteria (Audretsch et al., 2012; Lee et al., 2001; Zhou et al., 2014). This particular drawback signals the questionable generalizability of conclusions achieved through this approach (Blair and Zinkhan, 2006; King and He, 2005). Face-to-face interviews with founders promise particularly insightful results but require substantial time efforts and resources for implementation. Additionally, such studies are inherently biased as founders have a subjective perception of events, causes, and results.

Although researchers came up with a number of viable theories and explanations using the abovementioned methodologies, the lack of high-quality data and a number of issues impose significant limitations on researchers' ability to discover more substantial patterns and connections between observable phenomena, and advance theoretical understanding of firm growth (Coad, 2007; McKelvie

---

Keeley, 1990). TBNVs are used to be studied for more than thirty years with particular interest on the growth trajectories (Kazanjian, 1988; Lee and Lee, 2006; Strehle et al., 2010) and defining predictors of success (Reymen et al., 2017; Roure and Keeley, 1990; Shrader and Siegel, 2007; Symeonidou et al., 2017).

and Wiklund, 2010; Shepherd and Wiklund, 2009). For example, during the last sixty years, scholars are trying to explain and model the process of new venture development and to propose various organizational lifecycle models. According to the recent literature reviews, more than a hundred different models exist, of which all are conceptual (Levie and Lichtenstein, 2010; Salamzadeh, 2015). In other words, our understanding of new venture development is based on ideal constructs, lacking empirical validation, and data verification. Of course, conceptual models are valuable and useful in the number of cases, but at the same time, they have numerous limitations (or can even mislead!), especially when it comes to practical usage (Coad, 2007).

This research aims to tackle these issues and to contribute to overcoming the data scarcity problem in studying startups and high potential technology-based new ventures. We do so by demonstrating the credibility of web-search traffic information as a novel source of high-quality data in analyzing growth trajectories of high potential technology-based new ventures (TBNVs) emerged from them. Relying on the growing evidence that aggregated Internet search query data can be very useful in predicting underlying social and economic trends (Choi and Varian, 2012; Duwe et al., 2018; Jun et al., 2018; Wu and Brynjolfsson, 2009), we analyze the relationship between companies' growth trajectories represented by search activity, on the one hand, and by valuations achieved through rounds of investment, on another. We use a diverse and transparently selected sample of 241 US-based TBNVs from a variety of industries. The sample includes b2b and b2c companies, "unicorns" and "non-unicorns," digital platforms and traditional products. Their valuation data are collected from two leading databases on startups and TBNVs – Crunchbase[1] and CB Insights[2]. The search activity is measured using Google Trends[3] (GT), widely applied big data instrument (Jun et al., 2018).

Our results suggest that TBNV's growth dynamics are positively correlated with its web search traffic across the sample. This correlation is stronger when a company is a) more successful (in terms of valuation achieved) – especially if it is a "unicorn"; b) consumer-oriented (i.e., b2c) and 3) a digital platform. In the second step, to understand better which TBNV's feature or combination of features (i.e., b2b vs. b2c, "unicorn" vs. "non-unicorn," and digital platform vs. traditional products) leads to achieving high positive correlation between the TBNV's growth dynamics and its web search traffic, we employ fuzzy-set Qualitative Comparative Analysis (fsQCA) on our data. The results suggest that being a "unicorn" is a sufficient condition for the high positive correlation between the TBNV's growth dynamics and its web search traffic. However, it is not a necessary condition. A combination of consumer and digital platform orientation (i.e., b2c digital platform companies) is leading to the same result.

This study makes two contributions. First, by demonstrating that changes in Google Trends data reflect TBNVs' growth dynamics well, we verify a new methodology for analyzing and researching the growth of recently formed growth-oriented companies. The proposed methodology opens numerous possibilities to revisit some of the existing dilemmas in the field that was progressing slower than expected in recent years (Gilbert et al., 2006; Shepherd and Wiklund, 2009), and to provide new insights into the "how" aspect of growth, which is necessary and fundamental question that needs to be better understood (McKelvie and Wiklund, 2010). In this way, we add to the extant literature on firm growth (Aldrich, 1990; Davila et al., 2003; Greiner, 1972; Kazanjian and Drazin, 1990; Penrose, 1952; Shane and Venkataraman, 2000). To the best of our knowledge, this study is the first to validate usability of Internet search query data as a methodology to research startup growth. With many advantages – being public, free, easy to collect, available from the first day of company existence, and almost for each company – GT data can serve as a base for building data-driven trajectories describing and even predicting its evolution and growth, especially if they are "unicorns" and b2c digital-platform companies. Thus, our contribution to one of the central themes in entrepreneurship research is, primarily, methodological.

Second, our study contributes to the growing literature on using Google Trends data (Chumnumpan and Shi, 2019; Duwe et al., 2018; France et al., 2020; Jun et al., 2018, 2014b, 2014a; Rogers, 2018), extending its field of application from established companies to technology-based new ventures (including startups) and from technology management research to entrepreneurship.

The remainder of the paper is organized as follows. In Section 2, we present the theoretical background of our study and derive hypotheses. Next, in Section 3, we present the research sample and the methodology applied. Section 4 reports the results of the study, while Section 5 discusses their significance. After that, in Section 6, we summarize the main conclusions, observe limitations, and propose directions for future studies.

## 2. Theoretical background and hypotheses

Growth is crucial for startups – while established firms grow to sustain viability, startups grow to obtain it (Gilbert et al., 2006). Startups grow in a non-linear fashion (Garnsey et al., 2006) and almost always organically (McKelvie et al., 2006). However, the variance of their growth rates is considerably greater than in the case of established firms (Gilbert et al., 2006).

New venture growth process is dominantly described by stage models of growth (Greiner, 1972; Kazanjian, 1988). However, the researchers are far from the agreement about what stages exist, how many of them, and what are the relationships between them (Aidin, 2015; Levie and Lichtenstein, 2010; Zupic and Giudici, 2017). The alternative models, including those with tipping points (Phelps et al., 2007) and dynamic states (Levie and Lichtenstein, 2010) are also far from (wide) acceptance (Zupic and Giudici, 2017). Thus, a question "how startups grow?" stays fundamental question that needs to be better understood (McKelvie and Wiklund, 2010).

One of major limitations in answering the "how" question is lack of the appropriate growth indicator to apply (Achtenhagen et al., 2010; Shepherd and Wiklund, 2009; Weinzimmer et al., 1998), especially in context of growth-oriented new ventures. Different indicators have been used to measure growth, including sales growth, employment growth, asset growth, and equity growth. Followed by employment growth (variation in the number of employees) and market share growth (variation in the controlled share of a market), sales growth (variation in sales expressed as value) is proxy measure most frequently used in the literature to measure the growth of new ventures (Colombo and Grilli, 2010; Davila et al., 2003; Gilbert et al., 2006). Logic behind this is that when sales grow, a venture's revenues grow as well as the venture's ability to reinvest into resource expansion or capability development (Gilbert et al., 2006). Also, sales growth is easy to translate across countries and industry contexts (Delmar et al., 2003) and closer reflects entrepreneurs' point of view (Zupic and Giudici, 2017). However, all these proxies, including sales growth, are subjected to particular limitations as well (Shepherd and Wiklund, 2009). For example, market share growth is the indirect measure, which is dependent on the industry dynamics (Davila et al., 2003), while sales to happen, a new firm has to have a product or service available to sell (Gilbert et al., 2006). Also, different measurement practices across companies may affect their interpretation significantly.

To resolve these issues and better reflect entrepreneurs point of view (Achtenhagen et al., 2010), scholars recently started using a company valuation, achieved through funding rounds, as a proxy growth measure (Chang, 2004; DeTienne, 2010; Gornall and Strebulaev, 2017; Malyy et al., 2019). In this approach, the growth trajectories of venture capital backed new ventures are reflected by valuation data (DeTienne, 2010). Although it is generally accepted as a good representation of a new venture evolution (Davila et al., 2003; Gornall and Strebulaev, 2017; Malyy et al., 2019; Ratzinger et al., 2018), this approach has the same significant limitation as earlier mentioned sales and employee growth (and many other approaches) – data are rarely disclosed and available and for an outside observer, it is almost impossible to get enough objective information on a particular new venture's progress until it becomes public (i.e., carries out IPO) that is, however, also a relatively rare case.

## 2.1. Valuation and growth dynamics

Two basic premises of this research are that changes in market valuations of venture capital (VC) backed TBNVs reflect their growth dynamics and that statistics on web search activities is a good predictor of product acceptance by individuals or the society. Based on them, we argue that search traffic information, in particular, Google Trends data, will reflect well the growth dynamics of TBNVs. Building on this argument, we develop hypotheses relating growth dynamics of startups and TBNVs that emerge from them, with web search traffic.

High potential technology-based new ventures are understood to have various stages of growth (Kazanjian and Drazin, 1990; Lee and Lee, 2006; Tzabbar and Margolis, 2017), which condition the multiple series of VC investment – Pre-seed, Seed, Series A/B/C/etc., (Dahiya and Ray, 2012; Gompers, 1995) – and subsequent change in their market valuation (Davila et al., 2003). Financing through investment rounds is used to reduce information asymmetry and agency problems (Wang and Zhou, 2004). Each valuation event reflects investors' (re)assessment of a venture's ability to grow (i.e., generate future cash flows) and risks associated with that growth. To secure the first (and every other) funding round, TBNVs have to pass a systematic, disciplined and selective assessment. They have to provide VCs with the evidence about their high-growth potential, and progress in realizing that potential, by indicating elements like the attractiveness of the market, soundness of strategy, the feasibility of the technology, the existence of product-market fit, customer adoption, and the quality and experience of the management team (Davila et al., 2003). It was previously evidenced that venture capitalists are highly selective while making a decision to invest (Gompers and Lerner, 2001; Zider, 1998) and are known to be focused on the fast growth of the company's valuation (Zider, 1998) with simultaneous diminishing the potential risk of investment (Davila et al., 2003). Thus, during the early stages, the more convincing the evidence about the venture's progress is (e.g., the market is proved, and its size is significant, technology is feasible, schemes for value creation and value appropriation are verified, first customers are acquired, or a number of customers and sales are growing exponentially), the lower the risks and the higher valuation the venture will achieve. With each next funding and valuation round, new ventures are expected to show more tangible results reflecting their dynamics.

In this paper, we follow the logic of Davila et al. (2003). Namely, to examine the relevance of big data (i.e., Google Trends) for understanding the growth of new ventures, we study its relationship with changes in the value of equity. Equity valuation data for successive rounds of funding available from the CB Insights database allow us to estimate the growth of ventures over successive rounds (we selected only those with at least six investment rounds per venture to be sure sufficient data is present).

## 2.2. Web search and consumers' behavior

The big data taken from various sources have proven their value in management theory and practice. The studies discussing big data applications in terms of organization performance started to emerge in the early 2000s and gained strong momentum in 2010s (Batistič and der Laken, 2019). Considering the research subject, the related literature can be divided into two major clusters: focusing on the impact of big data on firms' performance and discussing new sources of big data for decision making. The studies from the first cluster, for instance, cover the application of big data for supply chain orientation (Gunasekaran et al., 2017; Wolfert et al., 2017; Yu et al., 2019), the role of big data in companies' capabilities context (Akter et al., 2016; Mikalef et al., 2019; Singh and El-Kassar, 2019), and predicting firm's manufacturing performance (Dubey et al., 2019; Ren et al., 2019). The works related to the second cluster discuss topics such as the use of customers' reviews in understanding sales performance (Lee and Bradlow, 2011; Mudambi and Schuff, 2010; Sheng et al., 2019), the influence of social networks users' information on business decisions (Antretter et al., 2019; Bradbury, 2011; Tambe, 2014), and application of web-search statistics for performance prediction and "nowcasting" (Choi and Varian, 2012; Jun et al., 2018; Shim et al., 2001; To et al., 2007). Since in the current research we use web search statistics as a source of big data, we will focus on reviewing this topic in more details.

The search feature of the internet was identified by scholars as one of the most important for users (Maignan and Lukas, 1997). Klein (1998) demonstrated that the process of search is specifically useful for obtaining information about goods due to the low costs of receiving objective data. This point was supported by another study (Liang and Huang, 1998), which claimed that effective minimization of transaction costs for the consumer is the key driver for the successful online selling of any product. Additional motivation for extensive search is the need to reduce the uncertainty about the product and decrease the level of the potential risk of purchasing something inappropriate (Dowling, 1986; Mitchell and Boustani, 1994). Thus, search over the internet during the pre-purchase process fits the basic economic theories and behavioral motives. Further studies demonstrated that the search over the internet and subsequent purchase under particular circumstances should be treated as the dependent processes due to the intention of consumers to use the same medium for getting data and obtaining goods (Shim et al., 2001; To et al., 2007).

Google Trends (originally known as Google Insights for Search[3]) is a big data tool. It was launched by Google in 2006. The first evidence of its usage for analyzing social trends came in 2009 when Google scientists Choi and Varian presented how Google Trends (GT) data could be applied for predicting automotive, retail, and home sales as well as for traveling (Choi and Varian, 2012). In the same year, GT data were used in a study published in Nature - Ginsberg et al. (2009) developed a model for detecting influenza epidemics using GT search query data that was consistently 1-2 weeks ahead of predictions of Centers for Disease Control and Prevention (CDC). After these pioneering studies, GT data were used in more than 700 studies across scientific fields, including business and management (Jun et al., 2018).

The willingness of consumers to decrease the pre-purchasing risk increases the importance of searching for information about it, especially for novel products (Assael, 1992). This phenomenon was demonstrated by Goel et al. (2010), who used GT search query volume to analyze and forecast the opening weekend box-office revenue for recent films, first-month sales for new video games, and the rank of songs in the Billboard Hot 100 chart. He showed that search statistics are generally predictive of consumer activities like attending movies, purchasing music, or video games that will happen in days or even weeks in the future. Considering the innovative products, web search queries can be extremely useful to analyze new technology adoption and able to provide higher explanatory power than indices used in the past, such as the GDP growth rate, patent applications, and news coverage (Jun et al., 2014b). It was also showed that brand search statistics explained the purchase of new technology adopters better than solely the number of search queries about the technology itself (Jun et al., 2014b). This fact gives us the ability to assume that web search of a TBNV brand name is a potentially strong predictor of future sales of this brand. In another study, Jun et al. (2014a) demonstrated that GT search traffic information related to a particularly innovative product serves as an accurate indicator of consumer attitudes towards it. In short, Google Trends data have a positive connection with actual consumers' interests in the novel products and new-technology brands and, by analyzing web search queries, it is possible to predict forthcoming sales, understand new technology levels of acceptance, and reveal apparent and hidden attitudes towards it. Thus, we hypothesize:

*H1: The TBNV's growth dynamics is positively correlated with web search traffic associated to it.*

## 2.3. TBNVs' features as the dimensions for analysis

### 2.3.1. "Success" dimension: "unicorn" vs. "non-unicorn"

A new venture becomes a "unicorn" when it reaches a valuation of $1 billion or more during at least one private series of venture funding (Gornall and Strebulaev, 2017). The value of "unicorns" is generally driven by VCs expectation that these TBNVs will significantly grow in the near future, becoming even more attractive for investment and, thus, generate above-average returns for the owners. These expectations are typically the consequence of investors' assessment (and perception) that *to-be-unicorns* are solving an important problem for a very large market in a way that is significantly faster, cheaper, safer or more comfortable (rarely combination of two or more of these features) than competing solutions, and can design a business model that will secure value appropriation and scale. Or, that they are solving an important problem for a very large market for the first time.

Valuing new ventures in different ways – some as "unicorns," some as "non-unicorns" – VCs, as informed agents (Baum and Silverman, 2004) offer us rare insight about growth predictions of these companies. Although the "unicorn" status does not guarantee future success (even survival), these ventures are considered as the most successful in their class. Reflecting this, for example, Fan (2016) proposes that "unicorns" should be regulated differently than "non-unicorns" due to the greater potential influence on the market of the former and increase of their investment risks. Thus, two poles in the success dimension are "non-unicorns" (successful companies) and "unicorns" (the most successful companies).

Many TBNVs generate only modest profits when they become "unicorns" but indicate significant future user interests and value for potential customers and, thus, the potential for growth. This interest could be first observed among early adopters and tech enthusiasts, then other users, who search the web for gaining more information about a new venture and its product (which typically is known under the same name). A growing number of studies show (Jun and Park, 2016; Shim et al., 2001; To et al., 2007) that the intent of web searches plays an important role in the intent to purchase a product, that is, to realize growth potential. This seems to be especially true when products are new (Assael, 1992; Jun and Park, 2016). Based on this argumentation, we expect the difference in how search traffic information will reflect (i.e., correlate with) the growth dynamics in the case of "unicorns" and "non-unicorns," and propose:

*H2: Growth dynamics of TBNVs with the "unicorn" status are better correlated with web search traffic associated to them then growth dynamics of TBNVs with the "non-unicorn" status.*

Achieving the "unicorn" status is not the only measure of a company's success. It is useful but binary, potentially hiding much of useful information. For example, it is not possible to distinguish between the effects of superior success (a "unicorn") and success (high-growing TBNV, but not a "unicorn"). A more granulated view on this can offer measures of relative success that show the speed of valuation growth. For that purpose, we adopt the approach of Ramadan et al. (2015) and use *Market Capitalization Growth Rate* (MCAP-GR), a measure that shows the annual growth rate of a company's market valuation. In the spirit of H2, we posit that the more successful TBNVs are, the better reflected their growth by web search traffic will be. However, this time we do not consider the company's success as a binary outcome, but through the prism of the speed of achieving success (i.e., maximum valuation), and hypothesize:

*H2a: The more successful a TBNV is, in terms of speed of valuation achieved, the better correlation will be between its growth dynamics and web search traffic associated to it.*

### 2.3.2. Customer type dimension: b2c vs. b2b

Further, we are interested in how the type of customers served by a TBNV influences the correlation between web search traffic and its growth dynamics. There are two main customer types – individual and business customers. They define two poles of the customer type dimension. Companies that make commercial transactions primarily with other companies are called business-to-business (b2b) oriented. Those who primarily serve individual customers are called business-to-customers (b2c) oriented. There are significant differences between these two types of companies (Ellis, 2010). B2b sales process is usually subjected to long negotiations (Järvinen et al., 2012), which are often focused on facts, i.e., terms of trade, reliability of delivery, etc. (Rėklaitis and Pilelienė, 2019). Decisions to buy from b2c companies are made faster, more emotionally, and more frequently (Saha et al., 2014). Further, b2b businesses are typically specialized and have a smaller number of customers (with bigger bills) than b2c businesses with the same cash flow. That means that in the case of b2c companies, much more independent purchase decisions happen, driving up information search activities of potential customers. In line with this and previous research that showed that search traffic correlates more strongly with sales of consumer goods than that of industrial goods (Jun and Park, 2016), we hypostatize:

*H3: Growth dynamics of b2c-oriented TBNVs are better correlated with web search traffic associated to them than the growth dynamics of b2b-oriented TBNVs.*

### 2.3.3. Product type dimension: digital platforms vs. traditional products

In this research, we distinguish between two types of digital products – individual (i.e., traditional) products and platforms[b] and use this differentiation to define two poles of this dimension. Platformization is a recent trend in developing products for a digitally connected world. It aims not at building stand-alone products but ecosystems that will profit from synergetic coexistence and collaboration of many complementary modules that are building their businesses on the platform (Nambisan et al., 2018). A number of sectors accepted digital platforms as an attractive business model and strategy (Asadullah et al., 2018). Typically, a platform provides the technological foundation (e.g., application programming interfaces and software development kits), legal protection (e.g., IP protection), and access to an established market (e.g., through platform's existing user base and reputation). All this with the goal to incentivize partnerships and development of complementary products and services, leverage economies of scale, and scope in innovation (Gawer, 2014). Further, digital platforms increase efficiency by significantly reducing costs of distribution, search, contracting, and monitoring (Asadullah et al., 2018). Successful platforms, like those in tourism (e.g., *TripAdvisor*[4] and *Expedia*[5]) or software development (e.g., *Apple iOS*, *Google Android*), are those that attract numerous complementors, which create value for platform users (Nambisan et al., 2018). Due to the synergetic effects of the created ecosystem, digital platforms lead to the increased number of platform users, with subsequent enhancement of the platform value (Asadullah et al., 2018). At the same time, the offers of multiple platform modules (complementors)

---

[b] In this study, digital platforms are defined as "two-sided networks that facilitate interactions between distinct but interdependent groups of users" (Asadullah et al., 2018), mediated by digital technology (Hein et al., 2019), not as the products, which consist of the functional core and the amount of possible third-party modules.

and the platform itself are expected to generate high information search activities of potential customers. Based on this, we hypothesize:

*H4: Growth dynamics of digital platform oriented TBNVs are better correlated with web search traffic associated to them then growth dynamics of traditional product oriented TBNVs.*

### 2.3.4. Configurational perspective

However, the three dimensions mentioned above do not exclude each other. New ventures are not only "unicorns" or b2b companies or platform owners. They may be all that at the same time – they can be "unicorn" b2b platforms or "non-unicorn" b2c traditional products or any of the other six possible combinations of these features. That means that there are eight different sub-groups of TBNVs, defined by a combination of the three features-dimensions ($2^3 = 8$). Although we predict (and aim to prove in this paper) that web search traffic is positively correlated with the growth dynamics of all high potential TBNVs, we also want to check if there are differences between sub-groups regarding this quality. We would like particularly to know if some TBNVs "always" lead to a high correlation with web search traffic while others do not. Or if there is a TBNV sub-group that has "always" low correlation.

We thus rely on configuration theory (Ketchen et al., 1997; Miller, 1986) to help us understand how the three dimensions combine to generate an outcome rather than how they individually compete to explain it. The fundamental proposition of the configuration theory is that outcome of interest can best be understood if patterns of causes are analyzed (Fiss, 2007; Misangyi et al., 2016). The specific causal patterns are called "configurations" (in our case, there are eight configurations, defined by different values of three features: company valuation, type of customer, and type of product). While emphasizing the importance of causal complexity, the configurational theory has three core assumptions (Misangyi et al., 2016): 1) there is rarely a single independent cause of an outcome and causes rarely operate in isolation from each other (multiple conjunction); 2) different configurations can lead to the same outcome (equifinality); and 3) the configurations that lead to the presence and absence of an outcome are not symmetrically opposite to each other (causal asymmetry);. Based on configurational theory and previous discussion, we hypothesize:

*H5: There are different configurations of TBNVs' features, i.e., the b2c vs. b2b, "unicorn" status vs. "non-unicorn," and digital platform vs. traditional product, that are equifinal in achieving high positive correlation between the TBNV's growth dynamics and web search traffic associated to it.*

## 3. Analysis and results

### 3.1. Description of the sample and data collection

In order to test the developed hypotheses, we used the popular startup databases: Crunchbase[1] and CB Insights[2]. The former provides a convenient search tool with the ability to use various filters and get the sample of companies meeting the general boundaries. We focused our study on TBNVs, which were founded in the US not earlier than on January 1, 2004, and not later than on August 31, 2019. These companies also should have reported on at least six rounds of venture financing, thus making it possible for us to obtain enough valuation points for analysis. The initial sample consisted of 2774 companies. For each company in the sample, we manually checked if it has enough (i.e., six and more) valuation points in the CB Insights database and collected valuation data for positive cases. We also excluded spinoffs of existing companies as search query data may be biased by a "mother" company noise and, thus, be unpredictably distorted. This stage of data collection resulted in 269 cases, for which we also aggregated their industry tags according to the CB insights classifications.

The last filter we applied accounted for the quality of the companies' GT data. Like any big data instrument, GT provides huge portions of information, which requires an assessment before usage. To assess the quality, we used a number of rules and developed a quantitative index (Appendix A). By applying them, we excluded 28 companies whose search query statistics were not good enough according to our index. For the rest of 241 companies, we automatically collected their GT data starting from the date of the company foundation and ending on August 31, 2019. For companies that we could not identify the exact date of foundation, we took January 1 of the corresponding year as the day of their foundation. Google Trends provides normalized values of search queries or, in other words, in each period of interest, there will be the point equal to 100 and other points related to it. In addition, to reach the weekly dimension of time series, only 200 points could be provided by GT what turns to the need to divide companies' lifecycles into 200-week periods (approx. 3.8 years). Taking into account the normalization of data built in the GT, after collection of the framed time series, we need to "sew" them together with the simultaneous repeating of normalization in order to have only one global maximum of 100 points.

After that, to eliminate the level of the fast noise, we also applied double exponential filtering (Huang et al., 2012; LaViola, 2003) for the valuation data with a small filtering coefficient: $\alpha = 0.99$. The coefficient $\alpha$ lies between zero and one, and values closer to the upper bound represent the lower level of filtration (Huang et al., 2012). During the last preprocessing step, in order to reach a larger number of analysis points, we applied linear interpolation to the valuation data, thus, making the number of examined points in GT and valuation data equal.

Finally, we obtained 241 research cases (Fig. 1) whose data were good enough to perform further correlation analysis. We have also analyzed the distribution of companies among industrial fields. According to the CB Insights principle of aggregation, there are three sequential levels of field identification from wider to more narrow[6]: industrial sector, industry, and sub-industry. The descriptive statistics of the sample are presented (Table 1). A more holistic representation of the distribution of the sample, according to the three selected dimensions, is offered in Table 2.

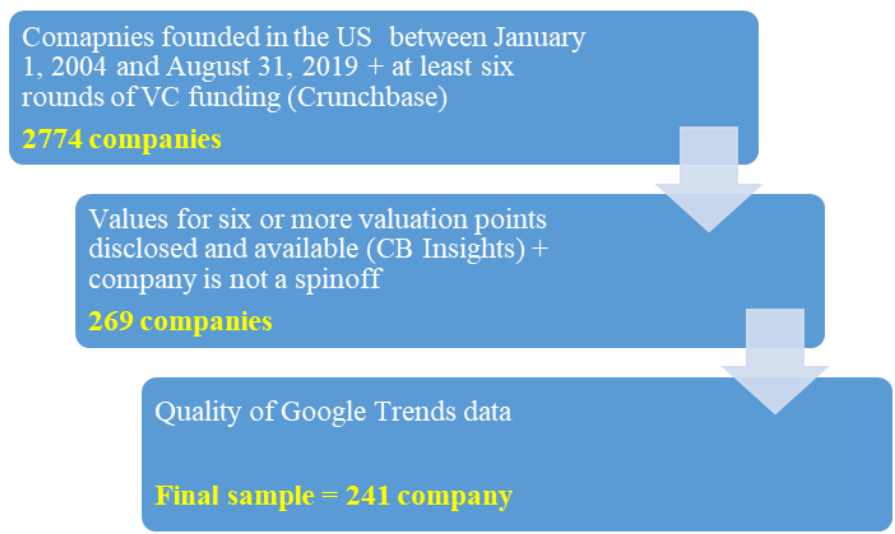

Figure 1. The sample selection and filtration process.

| Name of the feature | Number | in % of the sample |
|---|---|---|
| **Total** | 241 | 100% |
| **Success feature** | | |
| **Unicorns / Non-unicorns** | 106 / 135 | 44 / 56 |
| **Customer type feature** | | |
| **b2b / b2c** | 150 / 91 | 62 / 38 |
| **Product type feature** | | |
| **Digital platform / Traditional product** | 52 / 189 | 22 / 78 |
| **Top industrial sectors (1$^{st}$ level)** | 198 | 82% |
| **Internet** | 141 | 59% |
| **Mobile & Telecommunications** | 33 | 14% |
| **Healthcare** | 24 | 10% |
| **Top industries (2$^{nd}$ level)** | 168 | 70% |
| **Internet Software & Services** | 110 | 46% |
| **eCommerce** | 31 | 13% |
| **Mobile Software & Services** | 27 | 11% |
| **Top sub-industries (3$^{rd}$ level)** | 30 | 12% |
| **Business Intelligence, Analytics & Performance Mgmt** | 15 | 6% |
| **Advertising, Sales & Marketing** | 15 | 6% |
| **MCAP-GR, $M/year** | | |
| **Mean** | 373.60 | - |
| **25th pct** | 33.94 | - |
| **Median** | 92.11 | - |
| **75th pct** | 200.37 | - |

Table 1. Descriptive statistics of the sample.

| | "Unicorn" | | "Non-unicorn" | |
|---|---|---|---|---|
| | Digital platform | Traditional product | Digital platform | Traditional product |
| b2c | 28 | 23 | 14 | 26 |
| b2b | 6 | 49 | 5 | 90 |

Table 2. Holistic representation of the distribution of the sample.

## 3.2. Correlation analysis

### 3.2.1. Methodology

As a result of the data collection process, for each TBNV we obtained two series of data: search query statistics on the related term (i.e., company name) from Google Trends and the company's valuation data with interpolated points between the initial ones. Next, to decrease the level of noise while maintaining the weekly resolution of data, we applied the additional filtering to the GT time series with a stronger filtering coefficient $\alpha$. After a number of tests, we empirically tuned the value of $\alpha$ in such a way to provide the most similar curve to the original one, while keeping low amount of noise at the same time. To decrease the possible random errors, we have also applied low double exponential filtering to valuation data points (Fig. 2). Thus, for GT data, the filtering coefficient $\alpha$ equals *0.2*, while for the valuation time series, it equals *0.9*. After this step, we normalized the obtained search query and valuation data for the interval between 0 and 1 in order to receive the same scale of data series.

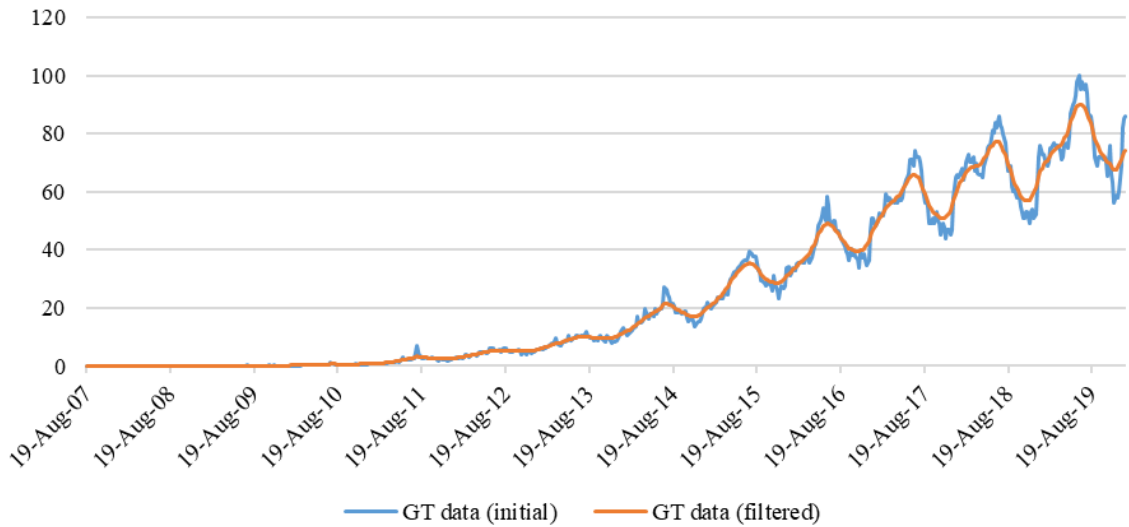

Figure 2. Example of the initial and filtered GT data ($\alpha$=*0.2*), case of *Airbnb*.

After finishing the data preparation step, we employed the rank correlation coefficient *Kendall's tau* (Kendall, 1975). Since (1) the *linear or non-linear* character of the assumed link remains unclear, (2) data *may not* be normally distributed, and (3) can be subjected to *outliers*, we selected the type of correlation measure, which provides the most reliable results in such uncertain scenarios (Allen, 2017). As for the threshold level, we selected 0.5[c], thus, treating the equal to and higher values as evidence of a strong link. Analysis of the level of Kendall's tau gives us the ability to make conclusions whether the character of search query data corresponds to dynamics demonstrated by a valuation history and in which manner.

---

[c] Kendall's tau is known to be connected with Pearson's rho by the following equation (Kendall, 1975): $R = sin(0.5*$ $Pi * tau)$. According to the established practice (Akoglu, 2018; Moore et al., 2012), a strong level of Pearson's rho is typically considered from 0.7. By applying backward calculations in the mentioned formula, we obtain that the strong level of Kendall's tau, thus, can be taken from 0.5.

Due to the reason that GT data of each case cover the whole lifecycle from the company foundation until the chosen date and series of investments are happening during the particular timespan of this lifecycle, the amount of search query data points is significantly larger than the number of valuation points. To deal with this fact, we adopt the approach from Kirk et al. (2013) – we presume that in some cases, a particular time lag between GT data and valuation dynamics may be present. Therefore, to check if the analyzed curves have similar dynamics but spaced in time, we employed the adopted cross-correlation method (ACC). The obtained shift in weeks, as well as the correlation coefficients, were recorded for each case, which did not provide a high correlation level otherwise. However, the method demonstrated high sensitivity that resulted in "false-positives". For instance, in some cases, the increase in correlation after applying the ACC method was minimal with, at the same time, relatively significant shifts between the series of data. In order to avoid such results, we measured the percentage of increase in Kendall's tau after applying the adopted cross-correlation and, conservatively, treated enhancements lower than 50% as not significant. Next, we present the obtained results.

### 3.2.2. Results of the correlational analysis

For each TBNV in the sample, we obtained a number of outputs, which included the highest Kendall's tau correlation level and the size of a weekly shift for the TBNVs when ACC was applied. The minus sign of the shift means that valuation data should be moved "back in time," i.e., growth in the GT data started earlier than in valuation. The opposite is correct as well: the plus sign tells that the valuation of the company started to grow before the GT search queries and should be shifted "forward in time" in order to reflect the highest correlation level. Considering the results, we can separate all sample TBNVs into three groups (Table 3):

1. G1 *"Strong link, without a shift"*: cases that showed a strong correlation and no shift was applied;
2. G2 *"Strong link, with a shift"*: cases that showed a strong correlation but a time shift exists;
3. G3 *"Weak link"*: cases that did not show a strong correlation in any option.

Examples of the plots with analyzed time series for all groups are presented (Fig. 3-5).

| | | | **Kendall's tau** | | | |
|---|---|---|---|---|---|---|
| | Count | Percentage | Mean | 25th pct | Median | 75th pct |
| **Total sample** | 241 | 100% | 0.66 | 0.56 | 0.70 | 0.82 |
| **G1 "Strong link, without a shift"** | 161 | 66.8% | 0.77 | 0.68 | 0.78 | 0.88 |
| **G2 "Strong link, with a shift"** | 39 | 16.2% | 0.63 | 0.56 | 0.63 | 0.70 |
| **G3 "Weak link"** | 41 | 17.0% | 0.25 | 0.15 | 0.38 | 0.45 |
| **Shift direction (from G2)** | | | | | | |
| **Positive shift** | 7 | 17.95% | 116.71 | 94 | 101 | 160 |
| **Negative shift** | 32 | 82.05% | -105.03 | -131 | -97 | -39 |

Table 3. Descriptive statistics of the obtained results.

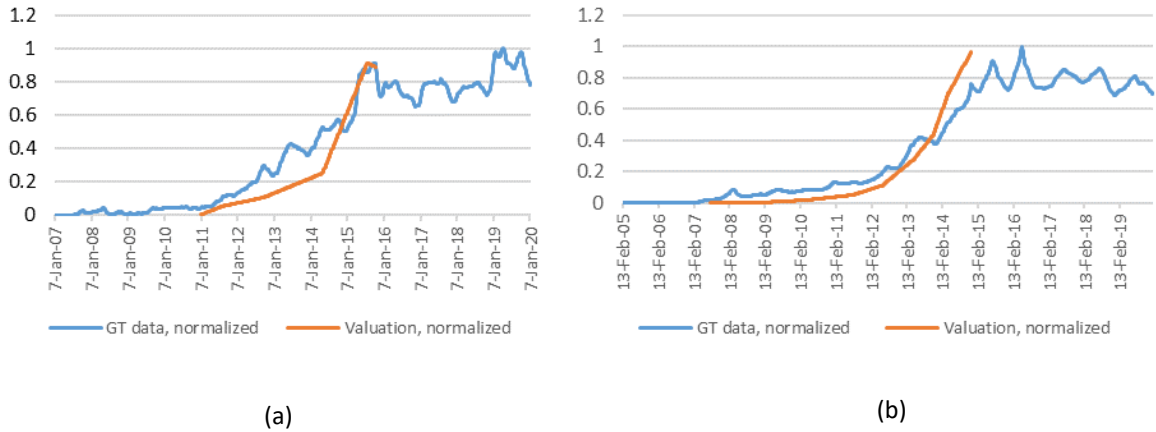

(a)                                        (b)

Figure 3. GT data and valuation curve of G1 "Strong link, without a shift": Kabbage (a) and Lending Club (b).

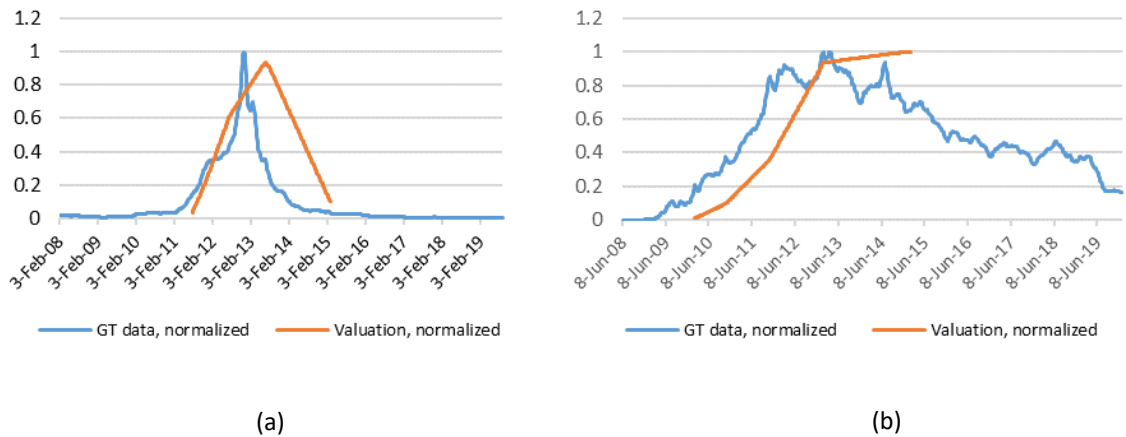

(a)                                        (b)

Figure 4. GT data and valuation curve of G2 "Strong link, with a shift": Fab (a) and Urban Airship (b).

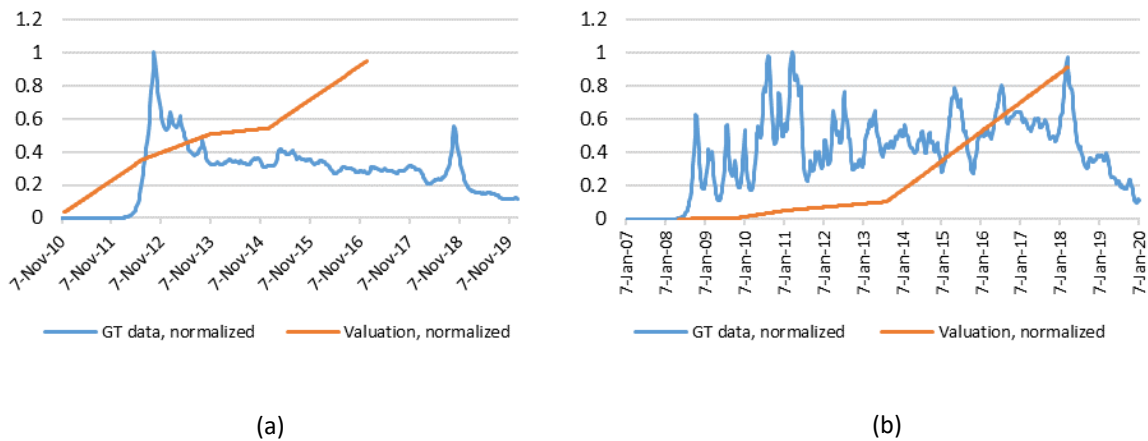

(a)                                        (b)

Figure 5. GT data and valuation curve of G3 "Weak link": Rethink Robotics (a) and NxThera (b).

Descriptive statistics of the obtained results demonstrate the left skew of Kendall's tau values for G1, "Strong link, without a shift," that signals stronger relationship than the correlation threshold chosen for the current research (Table 3). The cases with weak correlation have average Kendall's tau lower than 0.5 but still close to it, pointing at the existence of the particular level of the link between GT and valuation data but below the selected threshold. Among all cases with a time lag, absolute lag values are relatively significant: averages of both positive and negative variants are close to 100 what is approx. two years (Table 3, Positive shift, Negative shift). Considering the summarized results across the whole sample, it can be seen that the mean value of Kendall's tau equals 0.66, with 83% of the outcomes laying above the chosen threshold level with all p-values lower than 0.01 (Table 3, Total sample). The median value is also slightly higher than the mean that demonstrates the existence of the left skew among the obtained correlations. Percentiles also back this fact: the quarter of correlation coefficients in all cases from G1 lies above 0.88, which is significantly higher than the chosen threshold (Table 3, G1). Moreover, even the companies from G3, on average, obtain values quite close to 0.5, what signals about the existence of the particular relationship in weak correlation cases. Altogether, the obtained results present the clear evidence that TBNVs' dynamics are positively and significantly correlated with the related to them web search query statistics and, thus, support *H1*.

### 3.2.2.1. "Success" dimension

To describe the obtained results from the "success" dimension, we analyze how correlations between TBNVs' GT data and their valuations are distributed among the chosen groups. We can observe that 86% of all "unicorns" (92 cases) demonstrate a strong correlation without any time gap, as well as 51% (69 cases) of all "non-unicorns" (Table 4, Success dimension, G1 and G2). The fraction of all "unicorns," which show a high correlation is significantly higher than the fraction of all "non-unicorns."

At the same time, only 8% of all "unicorns" showed the presence of a time lag (G2), while for "non-unicorns," percentage is almost three times higher, 23% (Table 4, Success dimension, G2). The absolute average value of the time lag for "non-unicorns" is more than two times larger than for "unicorns" in the sample. The difference significantly decreases when median values are compared (87 vs 77 weeks). Smaller absolute time gap for "unicorns" belonging to G2 suggests better reflection of "unicorns'" growth dynamics in related to them web search queries (Table 4, Customer type dimension, Lag). For G3, the distribution between "unicorns" and "non-unicorns" is more or less the same as for G2: only a few "unicorns" (6%) do not show a strong correlation, while one quarter (26%) of all "non-unicorns" is in this group (Table 4, Success dimension, G3).

It can be observed that in G1, the "unicorn" companies obtain on average higher mean and median correlation values than "non-unicorns," and altogether are closer to one (represented by the 25th and 75th percentiles). At the same time, "unicorns" from G3 obtain lower median, while "non-unicorns" have more variations with some values falling below zero. This evidence suggests that the growth dynamics of TBNVs with the "unicorn" status are better reflected by their GT data, than the growth dynamics of TBNVs with the "non-unicorn" status, proving our hypothesis H2.

Next, we analyze companies' *Market Capitalization Growth Rate* (MCAP-GR) – another metric related to a company valuation. We took the maximum valuation of a TBNV, the date of this event, and then calculated how long it took a company to reach it, i.e., we obtained the maximum MCAP-GR. Results demonstrate that, on average, companies from Group 1 have a 37% higher MCAP-GR than the sample averages (Table 5, G1). In opposite, cases with weak correlations (G3) gain significantly lower average and median values compared to the whole sample: -69% for the average and -67% for the median (Table 5, G3). TBNVs from Group 2 obtained significantly lower than average measures of MCAP-GR (Table 5, G2) as well. However, it is expected - most companies from G2 have a negative time lag, so their valuation started to grow after some period since their foundation what decreased their MCAP-GR. The overall conclusion is that the faster the company's value is growing, the better its growth dynamics is reflected by GT search query data. It supports hypothesis H2a.

Altogether, considering the "success" dimension, the evidence shows that, as predicted by H2 and H2a, the more successful company is, the better it is reflected by related to it GT data.

| | | | COUNT | PERCENTAGE | | | KENDALL'S TAU | | | |
|---|---|---|---|---|---|---|---|---|---|---|
| | | | | Sample | Dimension | Group | Mean | 25th pct | Median | 75th pct |
| SUCCESS DIMENSION | "Unicorn" | G1 | 92/106 | 38% | 86% | 57% | 0.80 | 0.71 | 0.82 | 0.90 |
| | | G2 | 8/106 | 3% | 8% | 21% | 0.63 | 0.58 | 0.59 | 0.66 |
| | | Lag | - | - | - | - | -35 | -110 | -77 | 4 |
| | | G3 | 6/106 | 2% | 6% | 15% | 0.26 | 0.17 | 0.27 | 0.39 |
| | "Non-unicorn" | G1 | 69/135 | 29% | 51% | 43% | 0.72 | 0.64 | 0.73 | 0.80 |
| | | G2 | 31/135 | 13% | 23% | 79% | 0.64 | 0.56 | 0.62 | 0.70 |
| | | Lag | - | - | - | - | -73 | -121 | -88 | -46 |
| | | G3 | 35/135 | 15% | 26% | 85% | 0.25 | 0.15 | 0.38 | 0.45 |
| CUSTOMER TYPE DIMENSION | b2c | G1 | 72/91 | 30% | 79% | 45% | 0.78 | 0.68 | 0.80 | 0.89 |
| | | G2 | 7/91 | 3% | 8% | 18% | 0.70 | 0.61 | 0.66 | 0.74 |
| | | Lag | - | - | - | - | -37 | -99 | -54 | 34 |
| | | G3 | 12/91 | 5% | 13% | 29% | 0.28 | 0.25 | 0.43 | 0.46 |
| | b2b | G1 | 89/150 | 37% | 59% | 55% | 0.75 | 0.68 | 0.75 | 0.85 |
| | | G2 | 32/150 | 13% | 21% | 82% | 0.62 | 0.55 | 0.59 | 0.70 |
| | | Lag | - | - | - | - | -71 | -117 | -91 | -49 |
| | | G3 | 29/150 | 12% | 19% | 71% | 0.24 | 0.15 | 0.36 | 0.42 |
| PRODUCT TYPE DIMENSION | Digital platform | G1 | 40/53 | 17% | 76% | 25% | 0.83 | 0.77 | 0.87 | 0.90 |
| | | G2 | 6/53 | 2% | 11% | 15% | 0.70 | 0.60 | 0.70 | 0.77 |
| | | Lag | - | - | - | - | -42 | -100 | -62 | -37 |
| | | G3 | 7/53 | 3% | 13% | 17% | 0.30 | 0.21 | 0.40 | 0.42 |
| | Traditional product | G1 | 121/188 | 50% | 64% | 75% | 0.75 | 0.67 | 0.76 | 0.82 |
| | | G2 | 33/188 | 14% | 18% | 85% | 0.62 | 0.56 | 0.59 | 0.69 |
| | | Lag | - | - | - | - | -70 | -130 | -90 | -42 |
| | | G3 | 34/188 | 14% | 18% | 83% | 0.24 | 0.15 | 0.37 | 0.45 |

Table 4. Correlation analysis across three dimensions.

### 3.2.2.2. Customer type dimension

The distribution of results demonstrates that TBNVs focused on the b2c segment of the market have higher mean and median correlation values than b2b-oriented companies across all three groups.

**MAX MCAP-GR, $M/YEAR**

| | Mean | deviation from the sample, % | 25th pct | deviation from the sample, % | Median | deviation from the sample, % | 75th pct | deviation from the sample, % |
|---|---|---|---|---|---|---|---|---|
| **SAMPLE** | 373.60 | | 33.94 | | 92.11 | | 200.37 | |
| **G1** | 512.50 | 37% | 66.86 | 97% | 133.75 | 45% | 296.27 | 48% |
| **G2** | 71.59 | -81% | 18.51 | -45% | 37.44 | -59% | 117.70 | -41% |
| **G3** | 115.41 | -69% | 15.38 | -55% | 30.55 | -67% | 62.72 | -69% |

Table 5. Distribution of the *Market Capitalization Growth Rate* measure among the groups.

Percentiles (25th and 75th) also suggest that the distribution of correlation values for b2c TBNVs, across all three groups is more skewed to the left than in the case of b2b companies (Table 4, Customer type dimension, G1, G2 and G3).

Further, almost 80% of all b2c companies from our sample demonstrate high a correlation without time lag, compared to only 60% of all b2b companies from the sample. At the same time, less frequently b2c companies show weak correlation compared with b2b companies (13% vs. 19%). In the case when high correlations with and without time lag are considered together, again b2c companies dominate (87% vs. 80%).

When it comes to the time lag, only 8% of b2c companies achieve high correlation with it, opposed to 21% of b2b companies. The absolute average value of the lag for b2c companies is almost twice smaller than for b2b-oriented companies in the sample. This result suggests better connection between b2c companies' growth dynamics and related to them web search queries (Table 4, Customer type dimension, Lag).

These results present the clear evidence that b2c oriented companies have a better correlation between their growth dynamics and related to them web search queries. Hence, *H3* is supported. However, the difference between two "poles" in this dimension (b2c vs b2b) is smaller than what was observed for the "success" dimension. Thus, we can conclude that the customer type dimension is less impactful on the link between TBNVs' growth dynamics and related to them web search query data, than "success" dimension.

### 3.2.2.3. Product type dimension

Three-quarters of all TBNVs who create digital platforms in our sample (76%) demonstrate a high correlation between their GT and valuation data without any time lag, opposite to 64% of companies, which produce traditional products (Table 4, Product type dimension, G1). When high correlations with and without a time lag are considered together, the difference still exists but is significantly smaller (87% platforms vs. 82% traditional products). The higher percent of companies that market traditional products show a weak correlation between their GT and valuation data compared to platform-oriented companies (18% vs. 13%).

The descriptive statistics of G1 and G2 also indicate the higher correlation level in the case of digital platform products. On average, percentiles (25[th] and 75[th]), mean and median values of correlation coefficients for digital platforms are 10% higher than for the traditional products (Table 4, Product type dimension, G1 and G2). With the exception of the 75[th] percentile, the situation is the same for G3. That suggests that the distribution of correlation values across all three groups is more skewed to the left in the case of platforms than non-platforms (Table 4, Product type dimension, G1, G2 and G3).

A time lag between GT and valuation data leading to high correlation is identified in 11% of platform-oriented companies and 18% those developing traditional products. That is the smallest difference across three dimensions. The absolute average value of the lag is bigger for non-platforms, but this difference is also smaller than in the other two dimensions.

Summarizing, the results of product type dimension analysis support *H4* and demonstrate that growth dynamics of companies that develop their products in the form of a digital platform are better (and stronger) correlated with their web search traffic represented by GT data than companies, which create traditional products.

### 3.2.3. Industrial area perspective on the results

The last feature of the sample companies that we employed to examine the obtained results in current research is the company industrial area. Since CB Insights classification of new ventures' industrial sectors has three hierarchical levels (*sector – industry – sub-industry[6]*), we analyzed the distribution of results also separately. As a result, we can notice that for the *sectors* level, four areas present significantly higher results (Table 6). According to the mentioned previous studies, consumers tend to use the same channel for executing the process of search of information about the product and subsequent purchasing it (Shim et al., 2001; To et al., 2007). Thus, since the top three sectors in various manners may become mediating channels for a data acquisition process, TBNVs related to them are expected to have a stronger link between web search statistics and sales dynamics that directly influence their valuation.

Considering the second level of classification (*industry*), the top three areas go in line with common sense: people actively use the internet for selecting, searching, and buying goods, and especially mobile or desktop software. The last level – *sub-industry* – may not be very representative due to the many possible options, however, according to the obtained results, areas that focus on the b2b segment of the market overall obtain more high-correlation "hits" (Table 6). This outcome is quite notable since, as we mentioned in the previous chapter, being a b2b does not lead to the highest levels of correlation between the company's GT data and its valuation points. We can assume that new business products heavily utilize internet technologies, and b2b customers tend to use web search more often to choose the appropriate solution.

| Industrial sectors, the 1st level | | | |
|---|---|---|---|
| **Industrial designator** | **Count of high correlation cases** | **Count of all cases** | **Percentage** |
| **Computer Hardware & Services** | 8 | 11 | 73% |
| **Healthcare** | 13 | 24 | 54% |
| **Internet** | 120 | 141 | 85% |
| **Mobile & Telecommunications** | 26 | 33 | 79% |
| **Industries, the 2nd level** | | | |
| **eCommerce** | 29 | 31 | 94% |
| **Internet Software & Services** | 91 | 110 | 83% |
| **Medical Devices & Equipment** | 5 | 9 | 56% |
| **Mobile Software & Services** | 20 | 27 | 74% |
| **Sub-industries, the 3rd level** | | | |
| **Advertising, Sales & Marketing** | 12 | 15 | 80% |
| **Business Intelligence, Analytics & Performance Mgmt** | 14 | 15 | 93% |
| **Customer Relationship Management** | 6 | 8 | 75% |
| **Marketplace** | 7 | 8 | 88% |
| **Monitoring & Security** | 6 | 8 | 75% |
| **Payments** | 4 | 8 | 50% |
| **Social** | 5 | 6 | 83% |

Table 6. Results of industrial area dimension

## 3.3. Configurational analysis

### 3.3.1. Methodology

To better understand which TBNVs' features (i.e., b2c vs. b2b, "unicorn" vs. "non-unicorn," and digital platform vs. traditional products) lead to achieving high correlation between the TBNV's growth dynamics and related to it web search traffic, we employ fuzzy-set Qualitative Comparative Analysis (fsQCA) on our data sample. This is a set-theoretic, cross-case, and diversity-based research methodology that allows a holistic comparison of individual cases while identifying comprehensive configurations across the sample (Ragin, 2000).

By taking into account multiple conjunctural causations, causal equifinality and causal asymmetry of different conditions (i.e., TBNVs' features) in relation to the outcome of interest (i.e., high or low positive correlation between the TBNV's growth dynamics and its web search traffic), fsQCA allows us to deal with the extant causal complexity (Fiss, 2007; Misangyi et al., 2016) in this part of the data analysis. We assume that a TBNVs' feature may lead to the outcome of interest only in configuration with other features (i.e., the premise of multiple conjunctural causation), while different configurations of the TBNVs' features may be related to the same outcome (i.e., the premise of causal equifinality). Also, we assume that configurations of TBNVs' features related to the presence of the outcome of interest may not be symmetrical to configurations related to the absence of this outcome (i.e., the premise of causal asymmetry).

Thus, to be able to capture the complexity of relationships between the TBNVs' features and the outcome of interest, we conducted the fsQCA with the support of the algorithm of R Studio QCA package (Dușa, 2018) for the calibration of measures, analysis of necessity, as well as analysis of sufficiency. As fsQCA may treat varying degrees of case membership in sets, it allows the analysis of both fuzzy sets, from entirely out (value 0) to entirely in (value 1), and crisp sets, as entirely out (value 0) or entirely in (value 1).

Being based on the TBNVs' features, in our analysis the condition sets have the characteristics of crisp sets (i.e., b2c vs. b2b, "unicorn" vs. "non-unicorn," and digital platform vs. traditional products). Thus, after translating the qualitative data collected for the specific TBNV features into index measures for each of the condition sets, membership in the set is coded in the following way:

1. "1" for b2c-oriented venture and "0" for b2b oriented venture;

2. "1" for digital platform oriented venture and "0" for traditional products oriented venture; and

3. "1" for a venture with the "unicorn" status and "0" for a venture with the "non-unicorn" status.

Having the interest in the TBNVs' features leading to both high and low positive correlation between the TBNV's growth dynamics and related to it web search traffic, in this analysis we define two outcome sets, i.e., "high correlation" and "low correlation." As the outcome measure is based on the quantitative scale of correlation scores, the data required calibration for further fsQCA. Calibration is the transformation of raw numerical data into fuzzy-set membership scores that express the degree to which cases belong to a set (Schneider and Wagemann, 2012). Relying on the direct method of calibration (Ragin, 2008), we specified the very high correlation score of 0.9 as the qualitative anchor determining full membership (1) and very low correlation score of 0.1 as the qualitative anchor determining full non-membership (0) in the "high correlation" set. As the qualitative anchor determining a cross-over point (0.5) for membership in this outcome set, we use the correlation score of 0.499 to avoid the case ambiguity problem that can be caused by the use of the correlation score of 0.5 exactly. The membership in the "low correlation" set is coded as the negation of the correlation scores described above.

Based on the coded and calibrated data, we further conducted the analyses of necessity and sufficiency within the fsQCA. To identify the TBNVs' features and/or configurations of these features that are necessary for the outcome of interest, we use the consistency threshold of 0.9 and the relevance threshold of 0.6 in the analysis of necessity (Schneider and Wagemann, 2012). On the other hand, to identify the TBNVs' features and/or configurations of these features that are sufficient for the outcome of interest, we use the consistency threshold of 0.75 and the frequency threshold of 1 in the analysis of sufficiency, supported by the truth table analysis and logical minimization process (Schneider and Wagemann, 2012).

### 3.3.2. Results of the configurational analysis

The truth table analysis identifies eight different configurations of the three TBNVs' features (equals to $2^3$ possible configurations). The minimization process of the fsQCA identified two solutions that are sufficient for the high positive correlation between the TBNV's growth dynamics and related to it web search traffic, covering five different configurations of the TBNVs' features. There is no solution identified to have the sufficiency relation to the low correlation between the TBNV growth dynamics and related to it web search traffic (Table 7) – the three remaining configurations are identified to be insufficient for both high and low correlations between a TBNV's growth dynamics and related to it web search traffic. Also, there are no TBNV's features and/or configurations of these features identified to have a relevant necessity relation to neither of the two outcomes of interest.

| Solution | HIGH1 | HIGH2 | LOW |
|---|---|---|---|
| **Conditions:** | | | |
| b2c (●) vs. b2b (○) | - | ● | |
| Digital platform (●) vs. traditional product (○) | - | ● | No solution |
| "Unicorn" (●) vs. "non-unicorn" (○) | ● | - | |
| **Solution consistency and coverage:** | | | |
| Consistency | 0.82 | 0.81 | |
| Raw coverage | 0.50 | 0.20 | - |
| Unique coverage | 0.37 | 0.07 | |
| **Overall solution consistency and coverage:** | | | |
| Overall solution consistency | | 0.82 | |
| Overall solution coverage | | 0.57 | |
| No. of cases covered by the overall solution | | 120 | - |
| No. of cases not covered by the overall solution | | 121 | |

*Black circles represent core present conditions; dashes indicate conditions of indifference.*

Table 7. Results of fsQCA analysis.

The first solution (HIGH1) shows that a single TBNV's feature, i.e., the "unicorn" status, is a sufficient condition for a high correlation between its growth dynamics and the related web search traffic. The TBNVs' features of b2c or b2b and digital platform or traditional products appear as conditions of indifference. The solution HIGH1 shows high consistency of 0.82 and substantial coverage of 0.50.

Conversely, the second solution (HIGH2) shows that a combination of b2c and digital platform is sufficient for a high correlation between the TBNV's growth dynamics and related to it web search traffic. In this solution, the TBNV's feature of the "unicorn" or the "non-unicorn" status appears as a condition of indifference. The solution HIGH2 also shows high consistency of 0.81 and significantly lower coverage of 0.20 in comparison to the solution HIGH1.

The overall solution (HIGH1 + HIGH2) shows high consistency of 0.82 and substantial coverage of 0.57.

### 3.4.    Outlying cases

In the previous sections, we examined how various features of new ventures are connected with the correlation between their Google Trends search query data and valuation points. We demonstrated that the "unicorn" status, focus on the b2c customer segment, and being a digital platform leads to a higher correlation on average that was supported by a configurational analysis. The configurational analysis also demonstrated that to have a strong correlation between a TBNV's dynamics and related to it web search traffic, a company should be a "unicorn" or a b2c-oriented digital platform. Nevertheless, we obtained six "unicorns" (four of which are also b2c-oriented digital platforms) and one "non-unicorn" b2c digital platform that still demonstrate low correlation levels. Bearing in mind that TBNVs, which strongly "unfollow" our hypotheses, may provide additional valuable insights, next, we present an overview of these companies and discuss the possible reasons for their results.

In our outliers' list there are seven TBNVs: *Pinterest*[7], max tau = 0.28; *Compass*[8], max tau = 0.14; *Quantenna Communications*[9], max tau = 0.25; *Coinbase*[10], max tau = 0.43; *Coursera*[11], max tau = 0.03; *Marqeta*[12], max tau = 0.42; *TabbedOut*[13], max tau = 0.45. Five of these companies work in b2c market segment and develop digital platforms (*Pinterest, Compass, Coinbase, Coursera*, and *TabbedOut*). We could not derive a straightforward rule why the correlation between GT data and valuation points of companies is weak, but we assume that, for some cases, it may refer to the product and market positioning, while for others, the low correlation may be explained by limitations of the applied methodology.

For instance, the first group includes *Quantenna* Communications developing a discrete technology (Cohen et al., 2000), which is likely to have very low market value if considered separately from the complex product in that it is built-in. Another example is *Coinbase*, which created a digital currency exchange. Due to the novelty of the digital currency and high volatility of its most famous type (*Bitcoin*[14]), the service popularity might face a significant influence on the issues related to this topic[15,16]. The company's GT curve obtained a huge peak when digital currency started to bring the attention of a wide audience and dropped almost to the "pre-peak" values when the hype ended.

*Coursera* and *Pinterest* have similar graphs, thus, we try to find the same reason for their relatively low results. According to the data from open sources, the former TBNV corrected its business model several times[17] what might influence its perception in the eyes of investors and resulted in a slower speed of growth comparing to the dynamics of the GT curve. In its turn, *Pinterest* announced its first revenue-generating instrument *Promoted Pins* , late, five years after its launch[18]. In both cases, we can observe evidence of particular problems with value appropriation and, at the same time, successful value creation (Teece, 1986) that resulted in the fast growth of public interest in web search queries with significantly slower growth in valuation.

The second group of cases, whose low correlations may be justified by the methodology limitations, contain the rest of the outlying cases. It can be observed that GT data of *Marqeta* and *TabbedOut* face a significant level of noise, which should have been filtered out by a higher filtering coefficient. However, since we have selected the one level of filtering for all companies, we cannot manipulate it in specific cases after analysis of the results. In future studies, we plan to solve this issue by developing a case-dependent filtering algorithm. The last outlier, *Compass*, demonstrates a similar trajectory of GT data and growth in valuation, which is lagged for two hundred weeks. However, since in order to exclude a large number of false positives, we limited our maximum shift in the cross-correlation algorithm by the beginning of the valuation curve. Hence, in the case of *Compass,* we cannot reach the lag of two hundred points. This limitation is also planned to be eliminated in future studies. All plots for the outlying cases are presented (Fig. 6-9).

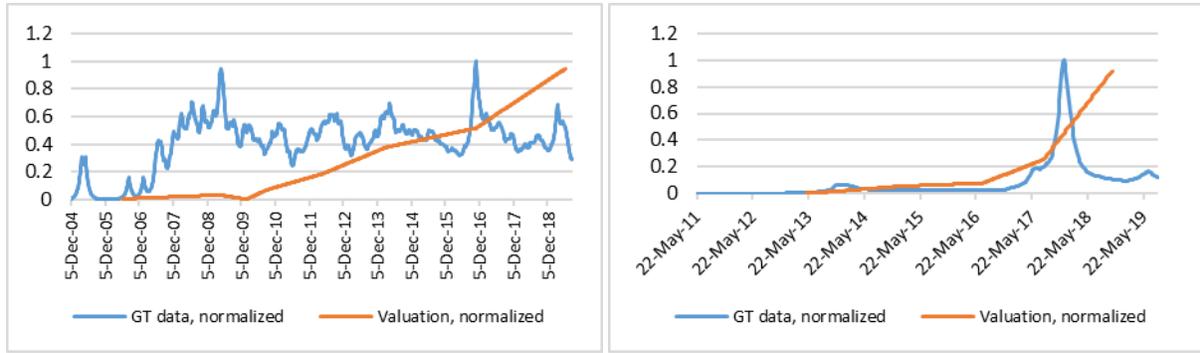

Figure 6. GT data and valuation curve of *Quantenna Communications* (a) and *Coinbase* (b).

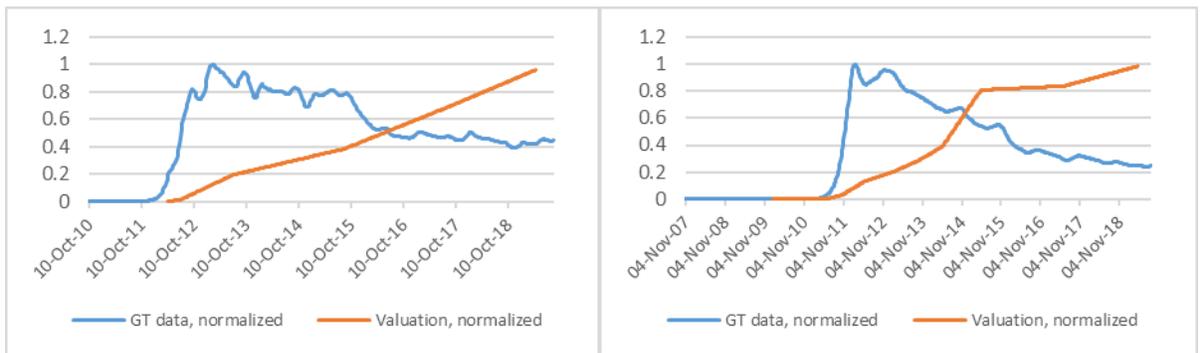

Figure 7. GT data and valuation curve of *Coursera* (a) and *Pinterest* (b).

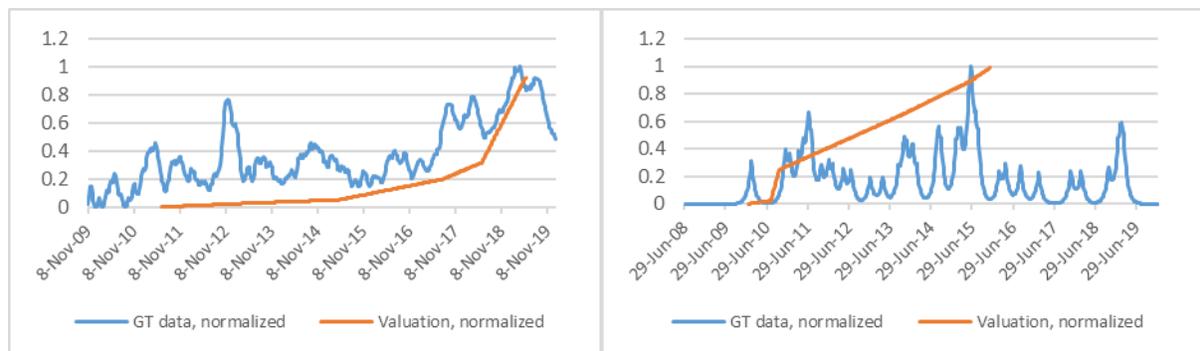

Figure 8. GT data and valuation curve of *Marqeta* (a) and *TabbedOut* (b).

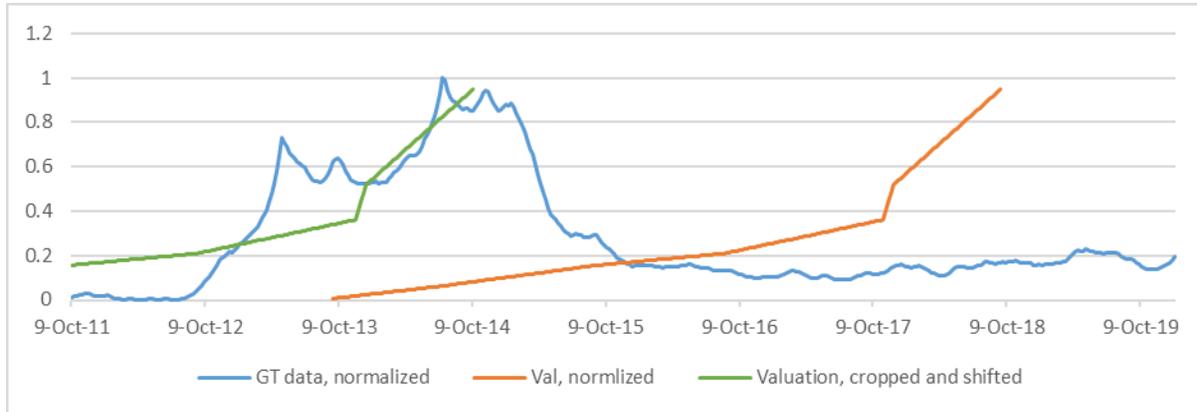

Figure 9. GT data and valuation curve of *Compass*.

## 4. Discussion

Across the diverse and large sample of TBNVs that we have analyzed in this paper, we observe that web search traffic generally correlates well with the companies' growth dynamics. Namely, majority of technology based new ventures from the sample, 66.8% of them, has correlation coefficient (Kendall's tau) between Google Trends search queries on their brand name and valuations they achieved through rounds of VC investments, higher than 0.5, without a lag, and statistically significant with all p-levels lower than 0.01. The additional 16.2% of companies from our sample have the same high correlation result when time shift between two sources of data is identified and taken into account through adopted cross-correlation. Altogether, 83% of companies from our sample show a high correlation. Moreover, the average value of the correlation coefficient across the whole sample is 0.66, with a median of 0.70; and the distribution of the correlation levels across the sample is heavily skewed to the left (Fig. 10).

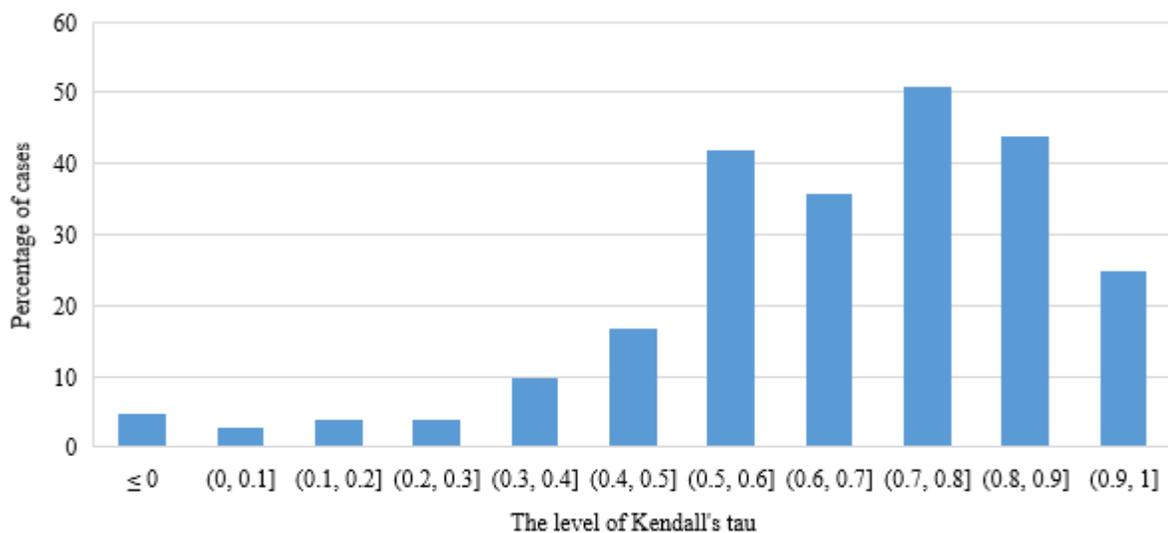

Figure 10. Distributions of Kendall's tau coefficient across the sample.

To figure out if the obtained correlations do not have spurious character, we implemented the test of the significance level for the obtained correlation coefficients for each case. Due to the fact that in all cases, we had more than ten points for correlational analysis (min = 107 points), we considered the obtained Kendall's tau as the normally distributed random variable (Kobzar, 2006, p. 625). Therefore, to make a conclusion with the chosen level of significance ($\alpha = 0.01$), we compared the obtained correlations with their normally distributed approximations, calculated as a product of the normal distribution α-quantile and the standard deviation of a related normal distribution (Kobzar, 2006). The results demonstrated that in 234 cases (or in 97% of the whole sample), the obtained correlation was not spurious with a 0.99 probability level. Taking to account the high average and median levels of correlation, we can conclude that the growth dynamics of TBNVs are positively and strongly correlated with the associated web search traffic.

When groups with the high correlation without a time lag (G1) and groups with the low correlation (G3) across the dimensions are analyzed and compared, we identify that for "unicorns" (success dimension) the difference is highest (86% vs. 6%), then for b2c companies (customer type dimension; 79% vs. 13%) and, then for digital platforms (product type dimension; 76% vs. 13%). In a similar manner, the success dimension exhibits the biggest difference between percentages of companies with the high correlation without a time lag (G1) at its different "poles" ("unicorn" vs. "non-unicorn": 86% vs. 51%), then customer type dimension (b2c vs. b2b: 79% vs. 59%) and, then product type dimension (digital platforms vs traditional products: 76% vs. 64%). However, when the strength of correlation achieved by the same companies (G1) is compared across the sample, the strongest correlation (highest mean and median values of correlation coefficients) with web search queries were observed for digital platforms (mean = 0.83, median = 0.87), then for "unicorns" (mean = 0.80; median = 0.82), and then for b2c companies (mean = 0.78; median = 0.80).

The results also demonstrate that 39 out of 241 companies in our sample exhibit a strong correlation between their growth dynamics and web search query data, once GT data points are shifted in time. A positive time lag signals that a company's value is growing faster than customers' interest in the product they are developing (as approximated by GT data), while the negative signals the opposite. Experience tells us that negative time lag should be much more frequent in startups' reality. That is reflected in our data – 32 companies display negative time lag, while only seven positive. Positive time lag may happen when VCs are especially enthusiastic about the team, product, or market due to the various reasons. On the other hand, the negative lag may be due to the problems with the value appropriation strategy, as it was observed in some outlying cases. In our sample, a time lag is bigger and more frequently identified in "non-unicorns," b2b-oriented companies and those, which market traditional products (Table 8). More precisely, in the success dimension, 8% of "unicorns" display a time lag, and 23% of "non-unicorns" (with the average time lag more than two times bigger for "non-unicorns"). In the customer type dimension, it is 8% of b2c companies vs. 21% of b2b companies with a time lag (with the average time lag almost two times bigger for b2b companies), while in the product type dimension, it is 11% of platform companies vs. 18% of non-platform companies (with the average time lag almost two times bigger for b2b companies).

| | "Unicorn" | | "Non-unicorn" | |
|---|---|---|---|---|
| | Digital platform | Traditional product | Digital platform | Traditional product |
| B2c | 1/28 | 0/23 | 3/14 | 3/26 |
| B2b | 1/6 | 6/49 | 1/5 | 24/90 |

Table 8. The distribution of cases with time lag in relation to all cases.

Overall, these results suggest that correlation between web search traffic and TBNVs growth dynamics is stronger for a) when ventures are more successful in attracting venture capital (H2a), especially for the most successful companies – "unicorns" (H2); b) when companies' customers are individuals (not the other businesses) (H3); and 3) when companies' products are in the form of digital platforms (not traditional products) (H4).

A better understanding of complex causal relationships between the three dimensions (or more precisely poles of these dimensions, i.e., "unicorn" vs. "non-unicorn," b2c vs. b2b, and digital platform vs. traditional products) and high correlation, comes after applying fuzzy-set Qualitative Comparative Analysis on our dataset. The results of the configurational analysis demonstrate that being a "unicorn" or a b2c-oriented digital platform are sufficient conditions that lead to a high correlation between web search traffic and TBNVs' growth dynamics, proving H5. Solutions determined by these two conditions cover, with high consistency (0.82) and relatively high coverage (0.57), five out of eight configurations in our analysis (Table 9). The three remaining configurations do not lead to low correlations (there are no configurations that lead to low correlation!). They are inconsistently related to both outcome sets, i.e., both high and low correlation between a TBNVs growth dynamics and web search traffic.

| | "Unicorn" | | "Non-unicorn" | |
|---|---|---|---|---|
| | Digital platform | Traditional product | Digital platform | Traditional product |
| b2c | | | | |
| b2b | | | | |

Table 9. Configurations of features, which lead to the high correlations between TBNVs' GT data and growth dynamics (*green*), and which are indifferent (*yellow*).

Our research makes two contributions. First, by demonstrating that changes in web search traffic reflect TBNVs' growth dynamics well, we verify a new methodology – a tool and data source – for analyzing and researching the growth of recently formed growth-oriented companies. In this way, we contribute to the extant literature on firm growth (Aldrich, 1990; Davila et al., 2003; Greiner, 1972; Kazanjian and Drazin, 1990; Penrose, 1952; Shane and Venkataraman, 2000). Being one of the key topics in the entrepreneurship (and management) literature, growth research has been attracting continuous and significant interest, but achieved only limited progress in recent years (Gilbert et al., 2006; McKelvie and Wiklund, 2010; Shepherd and Wiklund, 2009). By focusing on empirical and quantitative analysis (as recommended by Coad, 2007 and Achtenhagen et al., 2010) and verifying new indicator (as recommended by Weinzimmer et al., 1998) derived from open data, we validate new methodology allowing new insights into the "how" aspect of growth. This is a necessary and fundamental question that needs to be better understood to move the field forward (McKelvie and Wiklund, 2010). Our results reveal the potential of Google Trends data to be used as proxy measure of growth instead of non-public and rarely available measures like sales, employee and market share growth. Overcoming the limitations of the existing approaches, Google Trends data—which are public, free, easy to collect, available from the first day of company existence, and almost for each company — can help in building data-driven trajectories that will more accurately and, even, *in real time* reflect TBNVs' growth paths. These evolution curves should make it possible to revisit some old answers as well as to ask new questions and to come up with more solid concepts, theories, and predictions. That is especially true in the case of "unicorns" and b2c platform companies, but relevant even in the case of all other high-growth TBNVs, in case of which search traffic can still be useful for analyzing their growth dynamics albeit to a more limited degree.

Second, this is a pioneering study to use Google Trends data – big data created from human interaction with the Internet – to analyze startups and high-potential ventures emerging from them. Hence, we contribute to the recent literature using Google Trends data in business research (Chumnumpan and Shi, 2019; France et al., 2020; Jun et al., 2018, 2014b, 2014a). Several previous studies showed that web search traffic information could help in understanding the adoption of technology or the purchase of a product (Chumnumpan and Shi, 2019; Goel et al., 2010; Jun et al., 2014a, 2014b; Jun and Park, 2016). However, all these studies used established (mostly large and well-known) companies and their, more or less, known products. Unlike established firms, which know what they do, for whom, who pays for it and how much, how a solution is delivered, how money is collected, and have an internal organization that serves all these activities, TBNVs during their startup phases are searching to answer all these questions and, thus, behave differently in many aspects (Blank, 2013).Thus, our study extends previous applications of Google Trends data from established companies to technology-based new ventures (including their startup stages) and from technology management to entrepreneurship research.

## 5.    Conclusions, limitations, and implications

Startups and high-potential technology-based new ventures are "black boxes." They share only a limited amount of data – those they want people to see and have time to make public. This fact makes it hard to study startups. Academic researchers and analysts from venture funds and policy-making bodies use different approaches to connect the pieces of data to explain and predict real-life events. Some of the attempts resulted in successful empirical methods and some in viable theories – but with ample of space for improvements. In this study, based on a diverse sample of 241 US-based TBNVs from a variety of industries, we demonstrate that web-search traffic information, in particular Google Trends data, can serve as a powerful source of high-quality data for analyzing growth trajectories of high potential technology-based new ventures emerged from startups. The results suggest that for the most successful companies ("unicorns") and consumer-oriented digital platforms (i.e., b2c digital platform companies) proposed approach may become what X-ray chamber is for studying the human body – cheap, easy, and non-invasive way to understand what is going on inside a technology-based new venture.

However, our research is not without limitations. First, our sample is US-centered, and, thus, our results should be carefully generalized in other regions, especially in China. However, this selection was intentional as the US is the world-leading market for successful TBNVs and VCs, and, at the same time, Google is the dominant search engine. Second, "unicorns" are over-represented in our sample (44%). Although this may influence some of our results, such an amount of "unicorns" was obtained "organically" during the companies' selection process and according to the rules described. Since "unicorns" are more expected to attract several series of funding during their lifecycles, it was more likely to find for them enough valuation data points needed for analysis. Further, our study is based on two valuable sources of data – Google Trends data and companies' valuation information from the Crunchbase and CB Insights databases, both of which impose some limitations. Google Trends provides processed rather than raw data points, so our results depend on unknown processing methodology. However, it is the same across the sample, so we see this issue to be not too influential. The level of noise in some of the GT data was too high, so the level of filtration we chose was not enough to avoid noise and detect the main trend. Thus, in future studies, it may be useful to develop a filtering method, which will depend on the parameters of data and vary among cases. Finally, a great majority of companies do not make their valuation data public. Although Crunchbase and CB Insights provide deep insight, it is very hard to verify data objectively. Therefore, we cannot exclude the fact that some valuation points were provided with an error. However, since the source of data is similar for all cases, again, we believe it does not influence our results. We also noticed that the date of the company foundation also varies when taken from different sources. Although we triangulated data sources, we cannot undoubtedly claim that the starting date of analysis should not be earlier.

Despite the mentioned limitations, our analysis has been quite promising, offering an important methodological contribution. We believe that our research opens a wide range of possibilities for future applications in practice and academia. We see the opportunity for a wide application of Google Trends data as a proxy for analyzing technology-based new ventures' dynamics of development. For instance, TBNVs' GT data may be valuable for a better understanding of marketing strategies, business models, and intellectual property management practices used in technology-based new ventures, and their results. That especially may be the case in understanding, in practice frequently used term, like product-market fit or business model validation, which still lacks appropriate tools for a fuller explanation. We assume that this can help to predict future development of the early-stage ventures what, in turn, will positively influence the development of the entrepreneurship and innovation management areas.

Second, the methodology of using GT data for analyzing the growth dynamics of a particular venture can be slightly modified and applied for growth prediction purposes. Since GT data is very comprehensive (time series can be presented even in the minutes scale) and since we demonstrated its correlation with companies' valuation dynamics, we expect that it can serve as a basis for building company-related mathematical models of evolution and future growth. Whereas mathematical models work in both directions, we can infer that next to explaining historical data, they can also be used for predicting companies' future growth or decline. We aim to tackle this question in future studies.

Third, the methodology developed in this paper can be further improved and studied. Additional mathematical apparatus can be applied to improve achieved results. We assume that more complex statistical analyses may uncover more dependencies or study more dimensions (e.g., appropriability regimes and IP rights), enhancing the generalizability of our results.

Finally, the link found in the current research implies the positive correlation between two sets of data but does not tell anything about causal connection. Does the change in valuation cause stable growth in the public interest? Or, vice versa, the high amount of search queries leads to the rise in valuation? Or, maybe, these two processes reinforce each other? Answering these questions in future studies will lead to a deeper understanding of new ventures' evolution process and the premises of their success or failure.

Considering the implications for practice, our research adds value by verifying the additional source of data that venture capitalists may employ, next to existing sources, in the investment decision-making process. By proposing the objective source of data with a description of use, our study can provide meaningful benefits in identifying potential market leaders and decreasing the information asymmetry and risk. With further improvements, and increased ability to make more solid data-driven decisions, our methodology may even make venture capital not so *venture* anymore.

# Appendix A – Rules of Google Trends data quality assessment

The first thing we applied to was the GT search tag for a particular term. For instance, Google Trends may provide various search tags for a name of a company, e.g., "Company," "Corporation," "Solar energy company," "Software," "Website," etc. which influence the quality of statistical data. The company name may also reflect the common term (e.g., "Stripe") that influences the quality of GT data as well, so we considered it too. We have also developed the measure of noise in search query data following the simple logic: the closer search data to the beginning of the analyzed period, the lower should be the GT values. The reason behind this logic is connected to the naturally low public fame of a TBNV during the first period of its lifecycle and its future increase with company marketing. To measure this value, we calculated the ratio between the mean at the beginning of GT data and the overall mean – the lower the rate between means, the less systematic error in the case. The last measure we used reflects the amount of the related search queries: more related queries have the search term, the better are its GT data. All rules and assessment principles are exhibited below.

1. Selection of the company brand name. The brand name of the company may be corrected during the search code selection process through adding the identifiers, e.g., "Inc.," "Corp.," ".com."

2. Selection of the starting point of a data time range as the company foundation date

3. Selection of the category related to the company title in the following priority:

   a. **Group A:**

      i. "Company," "Corporation"

      ii. Category describing particular type of company, for example, "Solar energy company," "Transportation company," "Photovoltaics company"

      iii. "Software," "Website"

   b. **Group B:**

      i. Related to the particular geographical location, for example, "Corporate campus in Lexington, Massachusetts," "Health in Holladay, Utah," "Software company in San Francisco, California"

      ii. "Topic"

      iii. Without categorical relation

4. Assessment of the results according to the following empirically developed criteria:

   a. Brand name uniqueness. By a unique brand name, we understand used for it words or combination of words written together (without blanks) that rarely could be met in the normal language, e.g., "Lyft," "Airbnb," "Twitter." Assessment criteria:

      i. **Good case, 1 point** – **unique** name and identified category is from the **Group A**

      ii. **Fine case, 0.7 points** – **not unique** name and identified category is from the **Group A**

      iii. **Suspicious case, 0.3 points** – **unique** name and identified category is from the **Group B**

      iv. **Bad case, 0 points** – **not unique** name and identified category is from the **Group B**

   b. Systematic noise in the time series. The level of noise is calculated as the ratio of means: mean GT index of the first year is divided by the overall mean. Assessment principle:

      i. **Good case, 1 point** – ratio of means $\leq 0.5$

      ii. **Suspicious case, 0.5 points** – $0.5 <$ ratio of means $\leq 0.85$

  iii. **Bad case, 0 points** – ratio of means > 0.85

 c. Fast noise in the time series. This type of noise is described by sudden outbreaks and fast drops to zero. It is assessed by the level of the overall mean:

  i. **Good case, 1 point** – overall mean $\geq 4$

  ii. **Suspicious case, 0.5 points** – $2 \leq$ overall mean $\leq 4$

  iii. **Bad case, 0 points** – overall mean < 2.

 d. Amount of the related search queries. According to the GT mechanism, "users searching for your term also searched for these queries." Assessment principle:

  i. **Good case, 1 point** – related search queries $\geq 10$

  ii. **Suspicious case, 0.5 points** – $5 \leq$ related search queries < 10

  iii. **Bad case, 0 points** – related search queries <5

5. Total assessment criterion is calculated in two steps.

 a. Firstly, we calculates the average point between the points for *brand name uniqueness*, *related search queries*, and the *level of fast noise*.

 b. Secondly, we calculated the average between the obtained during the precious step result and the *level of systematic noise*. Since the cases with high systematic noise does not follow the common sense growth from lower to higher, we treated significance of this value equally to the average of all other criteria.

6. Finally, we marked each company according to the obtained grade:

 a. Cases with criterion **lower than 0.6** are marked as **bad**

 b. Cases with criterion **higher than and equal to 0.6** are marked as **good**

Bad cases are also re-checked with other related brand names, even with lower priority category. If the assessment criterion demonstrates improvement, it is selected for the case. If improvement was not reached, bad case is excluded from the sample.

## Numbered endnotes